# Determining the Critical Condition for Flow Transition in a Full-Developed Annulus Flow

Hua-Shu Dou,[1*] Boo Cheong Khoo [2] , Her Mann Tsai[1]

[1]Temasek Laboratories, National University of Singapore, Singapore 117508
[2]Department of Mechanical Engineering, National University of Singapore, Singapore 119260
**\*Email:** tsldh@nus.edu.sg; huashudou@yahoo.com

**Abstract**

Axial flow in an annulus between two concentric cylinders is commonly seen in various flow devices used in chemical processing industries and petroleum science and engineering. The flow state in the annulus strongly influences the performance of fluid transportation in the devices. Therefore, the determination of flow state which is laminar flow or turbulent flow is an important task to predict the performance of the flow devices. In previous works, we have proposed an energy gradient method for studying the flow instability and turbulent transition. In this method, it is shown that the flow instability and turbulent transition in wall bounded shear flows depend on the relative magnitude of the gradient of the total mechanical energy in transverse direction and the rate of loss of the total mechanical energy along the streamwise direction for a given imposed disturbance. For pipe and plane Poiseuille flows, it has been demonstrated that the transition to turbulence for these wall bounded parallel flows occurs at a consistent value of the energy gradient parameter (Kmax). In present study, the critical condition for turbulent transition in annulus flow is calculated with the energy gradient method for various radius ratios. The critical flow rate and critical Reynolds number are given for various radius ratios. Then, the analytical results are compared with the experiments in the literature. Finally, the implication of the result is discussed in terms of the drag reduction and mixing as well as heat transfer in practical industrial applications of various fluid delivery devices.

**Keywords:** Annulus flow; Turbulent transition; Energy gradient; Energy loss; Full-developed flow; Critical flow rate; Radius ratio.

**1. Introduction**

Annulus flow passage is widely used in aero-engines, turbomachinery, petroleum engineering, and various chemical industrial devices. The flow dynamics in the annulus has significant influence on the performance and flow efficiency of the devices. The flow in the annulus may be laminar or turbulent depending on the behaviour of the main flow and the disturbance level. Generally, the disturbance level in turbomachinery is very high so that the flow is likely to be turbulence in most circumstances. However, the flow may be kept laminar if the flow passage is carefully designed. The determination of the critical condition for the flow transition in annulus is therefore of great interest. As is well known, most flows in petroleum engineering and various chemical industrial devices are non-Newtonian and the flow are more complicated (Giiciiyener and Mehmetoglu, 1996; Guzel et al., 2009). At these circumstances, it is natural that we do the research on Newtonian fluid in annulus first and then the related results may be extended to non-Newtonian flows.

There are a lot of studies for the flow in the annulus between concentric cylinders in recent years. These research works mostly concentrate on the convective heat transfer problems (Yoo, 1998; Zarate et al., 2001; Kang et al., 2001;). The full developed turbulence flow was simulated using LES (large eddy simulation) method (Lee et al., 2004; Padilla and Silveira-Neto, 2008). Linear stability analysis has been used to study the instability and transition under the convective heat transfer (Dyko et al., 1999; Labonia ang Guj, 1998; Petrone et al., 2006; Adachi and Imai, 2007). The friction factor in the annulus under natural convective flow has been measured in (Lu and Wang, 2008). In these works, some works study only the laminar flow, and the others only deal with the turbulent flow. Very few studies are found to study the turbulent transition (Padilla and Silveira-Neto, 2008; Petrone et al, 2006), still they are only related to the transition under convective heat transfer and subjected to Rayleigh instability. Linear stability analysis was carried out for annulus inertial flow (Mott and Jeseph, 1968) and it only gave qualitative result about the stability boundary which has not been compared with the experiment. The results of linear stability analysis in (Mott and Jeseph, 1968) could not be related to the pipe Poiseuille flow, but the authors tried to relate annulus to the plane Poiseuille flow. In summary, it would be useful if a prediction method for turbulent transition under inertia role in various flows including annulus flow is provided.

Flow instability and turbulence transition under inertial role have been challenging topics for fluid dynamists for more than a century. This problem has not been solved fully owing to many difficulties encountered (Schmid and Henningson, 2000; Drazin and Reid, 2004). There exist a few theories for the analysis of flow instabilities. These are the above mentioned linear stability analysis, energy method, weekly nonlinear method, secondary instability theory, and others (Schmid and Henningson, 2000; Drazin and Reid, 2004). Linear stability analysis was successful for some flows such as Taylor-Couette flow and Rayleigh-Benard flow. But, it is failed for inertial shear flows such as plane Poiseuille flow, pipe Poiseuille flow, and plane Couette flow (Schmid and Henningson, 2000; Drazin and Reid, 2004). The pipe Poiseuille flow (Hagen-Poiseuille) is stable by linear stability analysis for all the Reynolds number Re. However, experiments showed that the flow would become turbulence if Re (=$\rho UD/\mu$) exceeds a value of about 2000. Experiments also showed that disturbances in a laminar flow in the pipe could be carefully avoided or considerably reduced, the onset of turbulence was delayed to Reynolds numbers up to Re=O($10^5$). For Re above 2000, the characteristic of turbulent transition depends on the disturbance amplitude, below which transition from laminar to turbulent state does not occur regardless of the amplitude of the initial disturbance. Linear stability analysis of a plane Poiseuille flow gives a critical Reynolds number Re (=$\rho u_0 h/\mu$) of 5772, while experiments show that transition to turbulence occurs at the Reynolds number of order 1000 (Schmid and Henningson, 2000; Drazin and Reid, 2004). For plane Couette flow, the flow is stable for all Re from linear theory consideration, but the experiments show a critical value of Re of about 370.

The energy method uses the Reynolds-Orr equation and involves the integration of energy for the entire flow domain which indicates the balance between the production and the dissipation of the disturbance energy. Using the energy method, the critical Re is 49.6, 81.5 and 20.7, respectively, for plane Poiseuille flow, pipe Poiseuille flow and plane Couette flow (Schmid and Henningson, 2000). Therefore, the non-uniformity in using this method to predict the flow transition is still imminent. On the other hand, the occurrence of instability is strictly a local behaviour in the flow field and the flow during the transition is intermittent. The first occurrence of the flow instability generally takes place in the most "dangerous" positions as seen in the formation of turbulence spot, the cylinder wake, and the dynamic stall on the airfoil with large attack angle. Hence, a method considering the local flow behaviour may be the correct approach. The weekly nonlinear method (Stuart, 1971) and the secondary instability theory (Bayly et al., 1988) seem to give better results than the above-mentioned other methods and can explain some phenomena; however there is still discrepancy with experiments.

Sengupta et al. (2003) used a method by considering the variation of redistribution of the total mechanical energy in the flow field to study the stability. The total pressure is decomposed into primary and disturbance components, then the temporal variation of disturbance of the total pressure is observed in the calculation to judge the instability. This method does not depend whether the primary flow is two or three-dimensional and the instability is finally determined by the kinematics of the flow field only. This method has been successfully used to explain the observed temporal instability phenomenon. What we learned from this study is that the evolution of the mean flow plays an important role in the instability of flows rather than just noticing the disturbance.

Recently, Dou (2006a, 2006b) proposed an *energy gradient method* to study flow instability and turbulent transition. Applying the principles of conservations of momentum and energy in the disturbed flows, it is found that transition to turbulence depends on the ratio of the gradient of the total mechanical energy in transverse direction and the energy loss due to viscosity in streamwise direction for a given disturbance. It is demonstrated that the plane and the pipe Poiseuille flows have a consistent critical condition at the transition condition determined by experiments. This method is also employed to study the flow instability in boundary layer flow (Dou and Khoo, 2009) and Taylor-Couette flow between concentric rotating cylinders (Dou et al., 2008), and the results demonstrated the validity of the model. In this paper, it is our intension that the proposed principle of energy gradient method is applied to the straight flow in an annulus. Then, the critical flow rate and critical

Reynolds number are calculated for various radius ratios. The result from the theoretical analysis is compared with available experiment. The implications of these results in engineering application are discussed.

## 2. Energy gradient method revisited

Dou (2006a; 2006b) proposed a mechanism with the aim to clarify the phenomenon of transition from laminar flow to turbulence for wall-bounded shear flows. In this mechanism, the whole flow field is treated as an energy field. It is thought that the gradient of total mechanical energy in the transverse direction of the main flow and the loss of the total mechanical energy from viscous friction in the streamwise direction dominate the instability phenomena and hence the flow transition for a given disturbance. It is suggested that the energy gradient in the transverse direction has the potential to amplify a velocity disturbance, while the viscous friction loss in the streamwise direction can resist and absorb this disturbance. The flow instability or the transition to turbulence depends on the relative magnitude of these two roles of energy gradient amplification and viscous friction damping of the initial disturbance. The analysis has obtained very good consistent agreement for the plane Poiseuille flow, the pipe Poiseuille flow, and the plane Couette flow at the critical condition (Dou, 2006a; Dou, 2006b). It is also demonstrated that an inflection point existence on the velocity profile is a sufficient condition, but not a necessary condition, for flow instability, for both inviscid and viscous flows. Dou (2006a, 2006b) carried out detailed and rigorous derivations from physics to give a solid foundation for this method, and explained recent modern experimental results about the scaling of the threshold of disturbance amplitude with the Reynolds number in the literature (Dou,2006a; Dou,2006b; Dou and Khoo, 2009; Dou et al., 2008). This method can be called as "*energy gradient method*." Here, we give a short discussion for a better understanding of the work presented in this study.

For a given base flow, the fluid particles may move oscillatory along the streamwise direction if they are subjected to a disturbance (Dou,2006a; Dou,2006b; Dou and Khoo, 2009; Dou et al., 2008). With the motion, the fluid particle may gain mechanical energy ($\Delta E$) via the disturbance, and simultaneously this particle may have mechanical energy loss ($\Delta H$) due to the fluid viscosity along the streamline direction. The analysis showed that the magnitudes of $\Delta E$ and $\Delta H$ determine the stability of the flow of fluid particles. For parallel flows, the relative magnitude of the total mechanical energy gained from the disturbance and the total mechanical energy lost due to viscous friction determines the disturbance amplification or decay. Thus, for a given flow, a stability criterion can be written as below for a half-period (Dou,2006a; Dou,2006b; Dou and Khoo, 2009; Dou et al., 2008),

$$F = \frac{\Delta E}{\Delta H} = \left(\frac{\partial E}{\partial n}\frac{2A}{\pi}\right) \bigg/ \left(\frac{\partial H}{\partial s}\frac{\pi}{\omega}u\right) = \frac{2}{\pi^2}K\frac{A\omega}{u} = \frac{2}{\pi^2}K\frac{v'_m}{u} < Const, \qquad (1a)$$

and

$$K = \frac{\partial E/\partial n}{\partial H/\partial s}. \qquad (1b)$$

Here, $F$ is a function of coordinates which expresses the ratio of the total mechanical energy gained in a half-period by the particle and the total mechanical energy lost due to viscosity in the half-period. $K$ is a dimensionless field variable (function) and expresses the ratio of transversal energy gradient and the rate of the energy loss along the streamline. $E = \frac{1}{2}\rho V^2$ is the kinetic energy per *unit volumetric fluid*, $s$ is along the streamwise direction and $n$ is along the transverse direction. $H$ is the loss of the total mechanical energy per *unit volumetric fluid* along the streamline for finite length. Further, $\rho$ is the fluid density, u is the streamwise velocity of main flow, A is the amplitude of the disturbance distance, $\omega$ is the frequency of the disturbance, and $v'_m = A\omega$ is the amplitude of the disturbance of velocity.

Since the magnitude of $K$ is proportional to the global Reynolds number ($\text{Re} = \rho UL/\mu$) for a given geometry (Dou,2006a; Dou,2006b; Dou and Khoo, 2009; Dou et al., 2008), the criterion of Eq.(1) can be written as,

$$\text{Re}\frac{v'_m}{U} < Const \quad \text{or} \quad (\frac{v'_m}{U})_c \sim (\text{Re})^{-1}. \qquad (2)$$

This scaling has been confirmed by careful experiments observed for pipe flow and boundary layer flow, and this result is in agreement with the asymptotic analysis of the Navier-Stokes equations (for $Re \to \infty$).

In term of Eqs.(1a) and (1b), the distribution of $K$ in the flow field and the property of disturbance may be the perfect means to describe the disturbance amplification or decay in the flow. According to this method, it can be found that the flow instability can first occur at the position of $K_{max}$, for given disturbance, which is construed to be the most "dangerous" position. Thus, for a given disturbance, the occurrence of instability depends on the magnitude of this dimensionless function $K$ and the critical condition is determined by the maximum value of $K$ in the flow. For a given flow geometry and fluid properties, when the maximum of $K$ in the flow field exceeds a critical value $K_c$, it is expected that instability can occur for a certain initial disturbance (Dou, 2006b). Turbulence transition is a local phenomenon in the earlier stage. For a given flow, $K_{max}$ is proportional to the global Reynolds number. A large value of $K_{max}$ has large ability to amplify the disturbance, and vice versa. The analysis has suggested that the transition to turbulence is due to the energy gradient of the base flow and the disturbance amplification (Dou, 2006a), rather than just the linear eigenvalue instability type. In linear stability analysis, an eigenvalue instability of linear equations requires the disturbance to be infinite small. In fact, finite disturbance is needed for the turbulence initiation in the range of finite Re as found in recent detailed experiments with injection as a disturbance source (Hof et al., 2003). Dou (2006b) showed that the disturbance amplitude for turbulent transition is scaled with Re by an exponent of -1, which is in agreement with the experimental data (Hof et al., 2003).

Dou (2006a; 2006b) demonstrated that the criterion obtained has a consistent value at the condition of transition determined by the experimental data for plane Poiseuille flow, pipe Poiseuille flow as well as plane Couette flow (see Table 1). From this table it can be deduced that the turbulence transition takes place at a consistent critical value of $K_c$ at about 385 for both the plane Poiseuille flow and pipe Poiseuille flow, and about 370 for plane Couette flow (Dou, 2006a). This may suggest that the flow transition in parallel flows takes place at a value of $K_c \approx 370\text{-}385$. The finding further suggests that the flow instability probably results from the action of energy gradients, and not to the eigenvalue instability of linear equations. The critical condition for flow instability as determined by linear stability analysis differs largely from the experimental data for all the three different types of flows, as shown in Table 1. For plane Poiseuille flow, both the two definitions of Reynolds number are given because different definitions are found in literature. This method is also applied to boundary layer flow on a flat plate (Dou and Khoo, 2009) and Taylor-Coutee flows between concentric rotating cylinders and it is proved to be valid (Dou et al., 2008).

For plane Poiseuille flow, this said position where $K_{max}> K_c$ should then be the most dangerous location for flow breakdown, which has been confirmed by Nishioka et al (1975)'s experiment. Nishioka et al's (1975) experiments for plane Poiseuille flow showed details of the flow breakdown. The measured instantaneous velocity distributions indicate that the first oscillation of the velocity occurs at y/h=0.50~0.62. In Dou (2006a), Dou's energy gradient method showed that the flow breakdown takes place first at the location of y/h=0.58, which accords with the experiments (Dou, 2006a).

For pipe flow, in a recent experiments on pipe flow, Nishi et al. (2008) showed that the oscillation of the velocity profile first started at r/R=0.53-0.73 in the transition process, which is in agreement with the prediction based on energy gradient method that r/R=0.58 is the most unstable position.

It should be pointed out that although the energy gradient method is not directly started from Navier-Stokes equations, it is not contradictable to Navier-Stokes equations. Since both the energy gradient method and the Navier-Stokes equations are based on Newton's second law, they are compatible each other.

## 3. Velocity distribution and Flow Rate in the annulus

The Navier-Stokes equation for the full-developed flow in an annulus (Fig.1) of an incompressible Newtonian fluid (neglecting gravity force) in the cylindrical coordinates (r,z) reduces

$$0 = -\frac{\partial p}{\partial z} + \mu \frac{1}{r}\frac{\partial}{\partial r}\left(r\frac{\partial u_z}{\partial r}\right) \qquad (3)$$

where ρ is the fluid density, $u_z$ is the axial velocity, p is the hydrodynamic pressure, and μ is the dynamic viscosity of the fluid.

Integrating the above equation and using the boundary condition at the walls of inner and outer cylinders, the velocity distribution along the radius can be obtained as (Papanastasiou, 1999)

$$u_z = -\frac{1}{4\mu}\frac{\partial p}{\partial z}R^2\left[1-\left(\frac{r}{R}\right)^2 + \frac{1-k^2}{\ln(1/k)}\ln\frac{r}{R}\right] = u_0\left[1-\left(\frac{r}{R}\right)^2 + b\ln\frac{r}{R}\right], \quad (4)$$

where $k=R_1/R_2$ is the radius ratio ($R_2=R$, $R_1=kR$), and $u_0 = -\frac{1}{4\mu}\frac{\partial p}{\partial z}R^2$. Here, $u_0$ is the velocity at the axis of the cylinder for the pipe Poiseuille flow and $b = (1-k^2)/\ln(1/k)$.

The flow rate can be integrated as (Papanastasiu, 1999)

$$Q = \int_{R_1}^{R_2} u_z 2\pi r dr = \pi R^2 \frac{u_0}{2}\left[(1-k^4) - \frac{(1-k^2)^2}{\ln(1/k)}\right]. \quad (5)$$

The average velocity in the annulus is

$$U = \frac{Q}{\pi R^2 - \pi(kR)^2} = \frac{u_0}{2}\left[(1+k^2) - \frac{(1-k^2)}{\ln(1/k)}\right]. \quad (6)$$

The Reynolds number based on average velocity is therefore

$$\text{Re} = \frac{U\rho 2(R-R_1)}{\mu} = \frac{\rho U 2R}{\mu}(1-k) = \frac{\rho u_0 2R}{2\mu}(1-k)\left[(1+k^2) - \frac{(1-k^2)}{\ln(1/k)}\right]. \quad (7)$$

4. Calculation of K in the annulus

The parameter K for the flow in an annulus is using Eq.(1)

$$K = \rho u_z \frac{du_z}{dr} / \mu \left( \frac{1}{r} \frac{du_z}{dr} + \frac{d^2 u_z}{dr^2} \right). \tag{8}$$

Using the velocity distribution expressed by Eq.(4), the following equations can be derived,

$$\frac{\partial u_z}{\partial r} = \frac{u_0}{R} \left[ -\frac{2r}{R} + \frac{b}{r/R} \right], \tag{9}$$

$$u_z \frac{\partial u_z}{\partial r} = \frac{u_0^2}{R} u^2{}_0 \left[ 1 - \left( \frac{r}{R} \right)^2 + b \ln \frac{r}{R} \right] \left[ -\frac{2r}{R} + \frac{b}{r/R} \right], \tag{10}$$

$$\frac{1}{r} \frac{\partial u_z}{\partial r} = \frac{u_0}{rR} \left[ -\frac{2r}{R} + \frac{b}{r/R} \right], \tag{11}$$

$$\frac{\partial^2 u_z}{\partial r^2} = \frac{u_0}{R^2} \left[ -2 - \frac{b}{(r/R)^2} \right], \tag{12}$$

$$\left( \frac{1}{r} \frac{du_z}{dr} + \frac{d^2 u_z}{dr^2} \right) = -\frac{4u_0}{R^2}. \tag{13}$$

Introducing above relations to Eq.(8) and simplifying, the following equation for the parameter K can be obtained,

$$K == \frac{1}{2} \frac{\rho u_0 R}{\mu} \left[ 1 - \left( \frac{r}{R} \right)^2 + b \ln \frac{r}{R} \right] \left[ \frac{r}{R} - \frac{b}{2(r/R)} \right]. \tag{14}$$

Let

$$\mathrm{Re}_1 = \frac{\rho u_0 R}{\mu} = \frac{\rho (u_0/2) 2R}{\mu} \tag{15}$$

be the Reynolds number of the virtual pipe flow, Eq.(14) can be further expressed as

$$K = \frac{1}{2}\text{Re}_1\left[1-\left(\frac{r}{R}\right)^2 + b\ln\frac{r}{R}\right]\left[\frac{r}{R} - \frac{b}{2(r/R)}\right]. \tag{16}$$

When b=0, the above equation simplifies to the pipe flow expression (Dou,2006a). Equation (16) can be also re-written as,

$$K = \frac{1}{2}\text{Re}_1 f\left(\frac{r}{R},k\right), \tag{17}$$

where

$$f\left(\frac{r}{R},k\right) = \left[1-\left(\frac{r}{R}\right)^2 + b\ln\frac{r}{R}\right]\left[\frac{r}{R} - \frac{b}{2(r/R)}\right]. \tag{18}$$

The distributions of $f(r/R,k)$ versus r/R for various k are shown in Fig.2. It is found that there is a maximum and a minimum of $f(r/R,k)$ along r/R for given k. The minimum $f_{\min}$ occurs at the inner cylinder and the maximum $f_{\max}$ occurs at the outer cylinder. The magnitude of $f_{\min}$ is large than that of $f_{\max}$ for each radius ratio. It is also found that both the magnitude of $f_{\max}$ and $f_{\min}$ decreases with the increase of k, and the both positions of $f_{\min}$ and $f_{\max}$ migrate to the outer cylinder with the increasing *k*. The maximum and minimum of the function *f* is listed in Table 2 for various radius ratio *k*.

It is found from Eq.(17) that there is a maximum or minimum of *K* for given $\text{Re}_1$, which corresponds to those of $f$, respectively. Since we concern the magnitude of the minimum or maximum of the function *f*, no matter how the sign of the minimum or maximum of the function of *f* for consideration of the flow instability, we take the following

$$f^*\left(\frac{r}{R},k\right)_{\max} = \max\left(\left|f\left(\frac{r}{R},k\right)_{\max}\right|, \left|f\left(\frac{r}{R},k\right)_{\min}\right|\right).$$

Then, we have the following equation from Eq.(17)

$$K_{\max} = \frac{1}{2}\text{Re}_1 f^*\left(\frac{r}{R},k\right)_{\max} = \frac{1}{2}\frac{\rho u_0 R}{\mu} f^*\left(\frac{r}{R},k\right)_{\max}. \tag{19}$$

It can be found from Table 2 that $K_{max}$ corresponds to those of the $f_{min}$ near the inner cylinder. If above equations are used to predict turbulent transition, the result means the onset of the turbulent transition on the side of inner cylinder.

**5. Comparison to Experiment of Critical Reynolds Number of the Annulus**

Walker et al. (1957) carried out some experiments for annulus flow. They employed the method of equivalent diameter to calculate the friction factor and then based on that to determine the critical Reynolds number for turbulent transition. Lohrenz and Kurata (1960) reviewed the experimental results for the calculation of friction factor for smooth circular conduits, concentric annuli, and parallel plates, annulus flow, and they found that there is a correlation of friction factor for these different flow passages. Hanks (1963) proposed a method based on the balance of the magnitude of the acceleration force and the magnitude of the viscous force to estimate the instability. The result of this method seems to be similar to the result of the energy gradient method. Later, Hanks and Bonner (1971) carried out further experiments for the annulus flow. Hanks's results have been used in (Giiciiyener and Mehmetoglu, 1996; Guzel et al., 2009).

From energy gradient method (Dou,2006a; Dou,2006b; Dou and Khoo, 2009; Dou et al., 2008), the dominating parameter determining the critical condition for instability is $K_{max}$. Dou (2006a; 2006b) demonstrated that the critical value ($K_c$) of $K_{max}$ at transition condition for wall bounded parallel flows (pressure driven flows) is a consistent quantity of about $K_c$ =385~389. Given this critical value of $K_c$ for the transition condition, the critical values of $u_0$, U, and Re can be calculated for any radius ratio of the annulus. As long as the values of these parameters are below the critical values, the flow in the annulus can be maintained as a laminar flow regardless of the level of disturbance. Obviously, this observation has significant implications in engineering.

At the critical condition, from Eq(19), we have the critical value of $u_0$ given as

$$u_{0c} = \frac{2\mu}{\rho R} \frac{K_c}{f^*_{max}}. \qquad (20)$$

Thus, at the critical condition, the critical flow rate is obtained from Eq.(5) as

$$Q_c = \pi R^2 \frac{1}{2} \frac{2\mu}{\rho R} \frac{K_c}{f^*_{max}} \left[ \left(1-k^4\right) - \frac{\left(1-k^2\right)^2}{\ln(1/k)} \right]. \tag{21}$$

The average velocity at critical flow rate is from Eq.(6),

$$U_c = \frac{Q}{\pi R^2 - \pi k R^2} = \frac{1}{2} \frac{2\mu}{\rho R} \frac{K_c}{f^*_{max}} \left[ \left(1+k^2\right) - \frac{\left(1-k^2\right)}{\ln(1/k)} \right]. \tag{22}$$

The critical Reynolds number is from Eq.(7),

$$\mathrm{Re}_c = \frac{U\rho 2(R-R_1)}{\mu} = \frac{\rho U 2R}{\mu}\left(1-\frac{R_1}{R_2}\right) = 2\frac{K_c}{f^*_{max}}\left(1-\frac{R_1}{R_2}\right)\left[\left(1+k^2\right)-\frac{\left(1-k^2\right)}{\ln(1/k)}\right]. \tag{23}$$

Using these equations, the critical flow rate $Q_c$ and the critical Reynolds number $\mathrm{Re}_c$ can be calculated for given geometry and value of $K_c$. The value of $f^*_{max}$ can be found from Eq.(18) (see Fig.2). Since the value of $K_c$ has been demonstrated to be a value of about 385~389 for Poiseuille flows according to experimental data and this value is assumed to be universal for wall-bounded parallel flows (pressure driven flows), and this value can be used to calculate the critical flow rate $Q_c$ and the critical Reynolds number $\mathrm{Re}_c$ for the annual flow.

According to the calculation in Fig.2, the magnitude of minimum of K on the inner cylinder is larger than the magnitude of maximum of K on the outer cylinder for a given Re. Thus, when the Re reaches a certain value, turbulent transition takes place first at the inner cylinder. After that, with further increase of Re, the maximum of K at the outer cylinder reaches another critical value, the turbulent transition then occurs at the outer cylinder. As such, the minimum critical Re for turbulent transition is determined by the lower one i.e., the flow at the side of the inner cylinder.

The critical condition of the theoretical result in this study is compared with experiments (Hanks and Bonner, 1971) and is shown in Fig.3 for the turbulent transition at the side of the inner cylinder. It is found that very good agreement is obtained between them. In the reference (Hanks and Bonner, 1971), turbulent transition is marked by a sudden increase of the drag coefficient with the variation of Re in the experiment.

Since the turbulent transition at the side of the outer cylinder occurs after the transition at the inner cylinder, in other words, the flow around the inner cylinder is already turbulent when the transition of outer cylinder occurs. Thus, from the transition of inner cylinder to the transition of the outer cylinder, the velocity in the annulus is a concurrent state of laminar and turbulent flow, which is different from the full smooth laminar flow in the whole annulus. Therefore, the prediction of transition at outer cylinder based on the full laminar flow in the annulus may be different from the experiments. As such, the comparison of the prediction to the experiment for the outer cylinder is not carried out.

The critical flow rate and critical Reynolds number for the transition at the inner cylinder versus the radius ratio are shown in Fig.3 and Fig.4, respectively, here Kc=385 is used according to the experimental data for pipe flow listed in Table 1. In Fig.4, the critical flow rate is normalized by the critical flow rate for pipe flow. It is found from Fig.3 that the critical value of the Re increases with the radius ratio (k) of the annulus. When the ratio k is less than a value of 0.18, the critical value of the Re is less than that of the Poiseuille flow of circular pipe ($Re_c$=2000). When the ratio k is larger than a value of 0.18, the critical value of the Re is larger than that of the Poiseuille flow of circular pipe ($Re_c$=2000). Therefore, we can say that the inner cylinder initiates instability for k<0.18 and enhance stability for k>0.18. Similarly, the critical value of the flow rate Q increases with the radius ratio (k) of the annulus. There exists also a value of k (k=0.12) at which the critical value of the Q of the annulus flow is equal to that of the pipe Poiseuille flow. The implication of theses results is that we can change the critical value of Re by inserting an inner cylinder at the center of a pipe and thus we can control the occurrence of turbulence. In this way, turbulence flow in a pipe can be caused to relaminarise in an annulus for k>0.18. This change may increase the averaged velocity in the passage, but the drag force is not increased significantly if k is not too large. This is because the friction coefficient for laminar flow is much lower than that for a turbulent flow. In other hand, we can also insert a thin cylinder in the center of a pipe (k<0.18) to reduce the critical Reynolds number to enhance the mixing or heat transfer.

It should be emphasized that when the radius of the inner cylinder tends to zero, the annulus shows a singularity. In this case, the annulus is not equivalent to a cylinder, as shown in Fig.3 and Fig.4.

When the radius ratio $k \to 1$, the radii of the annulus tends to be infinity, the limit of the annulus flow approaches the plane Posiseuille flow. Thus, the critical Re for plane Poiseuille flow can be calculated from the equation of annulus flow with $k \to 1$. If the critical Re for plane Poiseuille flow obtained from annulus flow calculation is in agreement with the experimental data of plane Posiseuille flow, it would be demonstrated that the idea and the method employed in this study is correct for the calculation of critical Re number of annulus flow. When we look at Eq. (23), it is found that is singular for k=1. At this case, the critical Re for plane Poiseuille flow can be extrapolated with k=1 from Fig.3. Using this method, we obtain the critical Re for plane Poiseuille flow, 2679, as shown in Fig.3. Then, we refer to the definition of Re in Eq.(23), and we obtain $2U\rho(R-R_1)/\mu$ =2679. In consideration of the width of the limiting plane Poiseuille flow L=$(R-R_1)$, the critical Re for plane Poiseuille flow obtained is $U\rho L/\mu$ =1340. This value accords well with the experimental data listed in Table 1 in which the critical Reynolds number is 1350. Therefore, this result confirms that the criterion proposed for annulus flow in this study is correct. This confirming route is summarized as follow: (1) obtain the value of Kc from experimental data of pipe flow, say Kc=385; (2) derive the formulation of Re for annulus flow; (3) put this value of Kc into the equation of annulus flow as a criterion; (4) calculate the critical Re for annulus flow and plot it in the figure; (5) obtain the critical Re of plane Poiseuille flow by extrapolating the data of annulus flow to k=1. As we have seen, the predicted critical Re for plane Poiseuille flow is in perfectly consistent with the experimental data.

As discussed above, the radius ratio of the inner to the outer cylinders has important influence on the critical condition of transition to turbulence. This result can be used to many industrial processes and fluid transportations. For the purpose of the demand industrial applications, we can design the device to satisfy the requirement to change the flow transition, such as reducing drag force (increasing the $Re_c$), and enhance (reducing $Re_c$) or weaken (increasing $Re_c$) the mixing and the heat transfer, etc.

# 6. Conclusions

As is well know, even for simple pipe Poiseuille flow, plane Poiseuille flow, and plane Couette flow, there is no exact theoretical method to predict the critical condition of turbulent transition nowadays. For annulus flow between concentric cylinders, it is much more difficult to derive an exact theoretical method to predict the turbulent transition.

In this study, the critical condition for the instability of full-developed laminar flow in an annulus is given following the "energy gradient method." The criterion for instability is based on the energy gradient method for parallel flow instability. It is shown that the critical flow rate and the critical Reynolds number for onset of turbulent transition increase with the radius ratio of the annulus. It reaches to the critical Re of circular pipe flow at the radius ratio of about k<0.12~0.18, where the radius of which is same as the outer radius of the annulus. It is clear that the inner cylinder creates instability for k<0.12~0.18; and enhance stability for k>0.12~0.18. Following this principle, the flow transition can be controlled by changing the radius ratio for a given outer radius. The idea can be employed in the design of fluid flow devices to control the flow status like drag reduction (by keeping to laminar flow) or increase the mixing of the fluid media and the heat transfer (by changing to turbulent flow).

The critical Re of plane Poiseuille flow obtained by extrapolating the data of annulus flow to k=1, is perfectly consistent with the experimental data. This result confirms that the energy gradient method is applicable for pipe Poiseuille flow, plane Poiseuille flow as well as annulus flow for turbulent transition. This research provides with the foundation in principle for the annulus flow of non-Newtonian fluid. The resulting this study can be extended to flows of non-Newtonian fluid flows following the principle of the energy gradient method.

| Flow type | Re expression | Eigenvalue analysis, $Re_c$ | Energy method $Re_c$ | Experiments, $Re_c$ | Energy gradient method, $K_{max}$ at $Re_c$ (from experiments), $\equiv K_c$ |
|---|---|---|---|---|---|
| Pipe Poiseuille | $Re = \rho UD/\mu$ | Stable for all Re | 81.5 | 2000 | 385 |
| Plane Poiseuille | $Re = \rho UL/\mu$ | 7696 | 68.7 | 1350 | 389 |
|  | $Re = \rho u_0 h/\mu$ | 5772 | 49.6 | 1012 | 389 |
| Plane Couette | $Re = \rho Uh/\mu$ | Stable for all Re | 20.7 | 370 | 370 |

Table 1 Comparison of the critical Reynolds number and the energy gradient parameter $K_{max}$ for plane Poiseuille flow and pipe Poiseuille flow as well as for plane Couette flow [19, 20]. $U$ is the averaged velocity, $u_0$ the velocity at the mid-plane of the channel, $D$ the diameter of the pipe, $h$ the half-width of the channel for plane Poiseuille flow (L=2h) and plane Couette flow. The experimental data for plane Poiseuille flow and pipe Poiseuille flow are taken from Patel and Head [24]. The experimental data for plane Couette flow is taken from Tillmark and Alfredsson [25], Daviaud et al. [26], and Malerud et al.[27]. Here, two Reynolds numbers are used since both definitions are employed in literature. The data of critical Reynolds number from energy method are taken from [1]. For Plane Poiseuille flow and pipe Poiseuille flow, the $K_{max}$ occurs at y/h=0.58, and r/R=0.58, respectively. For Plane Couette flow, the $K_{max}$ occurs at y/h=1.0.

Table 2 Maximum and minimum of function f.

| K | 0 | 0.1 | 0.2 | 0.4 | 0.6 | 0.8 |
|---|---|---|---|---|---|---|
| 2*fmax | 0.7698 | 0.2908 | 0.1941 | 0.07988 | 0.02382 | 0.003003 |
| 2*fmin |  | -0.4705 | -0.2678 | -0.09549 | -0.02625 | -0.003156 |

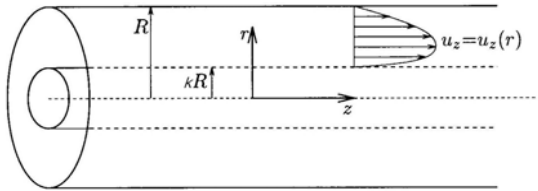

Fig.1 Sketch of the flow in an annulus. $R_2=R$; $R_1=kR$

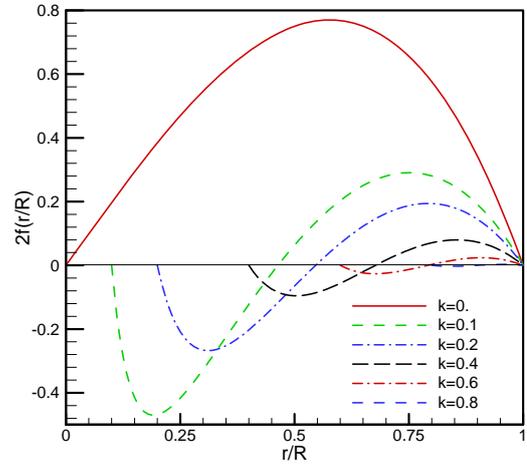

Fig.2 Function f(r/R) along the radius r/R for various radius ratios (k) between inner and outer radius. Outer radius of the annulus is the same.

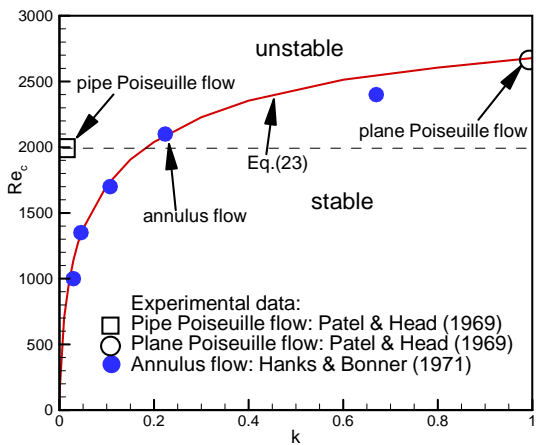

Fig.3 Comparison of prediction with experimental data for turbulent transition at the inner cylinder: Reynolds number at critical condition of inner cylinder for annulus flow versus the radius ratio k. The out radius is kept constant. The data at k=0 means the case for circle pipe flow. The data at k=1 corresponds to the case for plane Poiseuille flow.

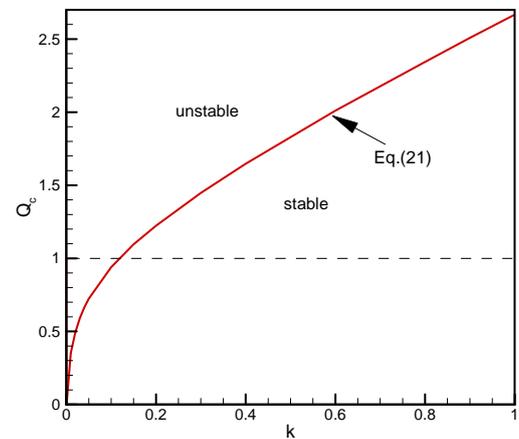

Fig.4 Flow rate at critical condition of inner cylinder for annulus flow versus the radius ratio k (Eq.(21)). The outer radius is kept constant. The data at k=0 means the case for circle pipe flow. The data at k=1 corresponds to the case for plane Poiseuille flow.